# Direct Imaging Strain-field Vortex Networks in Twisted Bilayer Graphene Magnified by Moiré Superlattices

Ya-Ning Ren, Yi-Wen Liu, Chao Yan, and Lin He*

Center for Advanced Quantum Studies, Department of Physics, Beijing Normal University, Beijing, 100875, People's Republic of China

*Correspondence and requests for materials should be addressed to Lin He (e-mail: helin@bnu.edu.cn).

**In two-dimensional (2D) twisted bilayers, the van der Waals (vdW) interlayer interaction introduces atomic-scale reconstruction at interface by locally rotating lattice to form strain-field vortex networks in their moiré superlattice. However, direct imaging the tiny local lattice rotation of the strain-field vortex requires extremely high spatial resolution and is an outstanding challenge in experiment. Here, a topmost small-period graphene moiré pattern is introduced to magnify sub-Angstrom distortions of the lattice and tiny local lattice rotation in underlying twisted bilayer graphene (TBG). The local periods and low-energy van Hove singularities of the topmost graphene moiré patterns are spatially modified by the atomic-scale reconstruction of the underlying TBG, thus enabling real-space imaging of the strain-field vortex networks. Our results indicate that structure-reconstructed vdW systems can provide a unique substrate to spatially engineer supported two-dimensional materials both in structures and electronic properties.**

Stacking two-dimensional (2D) van der Waals (vdW) bilayers with a controlled interlayer twist angle provides a new pathway to engineer the structures and electronic properties of the system with continuous tunability [1-5]. However, recent studies demonstrated that the structures of the 2D vdW twisted bilayers could be distinct from that with assuming a rigid rotation of the two adjacent layers [6-22]. In 2D vdW twisted bilayers, the vdW interlayer interaction, although weak in nature, can cause atomic-scale reconstruction at the vdW interfaces by favouring commensuration between the two adjacent layers, which competes with the intralayer elastic energy. Twisted bilayer graphene (TBG), as the most studied 2D vdW bilayers, exhibits substantial lattice reconstruction at the interface for small twist angle $\theta$, significantly changing its lattice symmetry and electronic structure [11,12,15,21,22]. The interfacial reconstruction in the TBG is achieved by rotating the lattice locally to form strain-field vortex networks in the moiré superlattice. Recent developments in state of the art microscopy techniques have enabled the characterization of atomic-scale reconstruction in the TBG [22], but direct imaging the tiny local lattice rotation of the strain-field vortex requires extremely high spatial resolution and has remained an outstanding challenge in experiment.

In this Letter, we demonstrate a general approach for direct imaging the local lattice rotation of the strain-field vortex in the small-angle TBG by using a topmost large-angle (small-period) graphene moiré. The sub-Angstrom distortions of the lattice and tiny local rotation in the underlying small-angle TBG are largely magnified (about two orders, see Fig. S1) by the topmost small-period graphene moiré, resulting in its spatial modified twist angles and low-energy van Hove singularities (VHSs). This enables us to direct visualize the tiny local lattice rotation of the strain-field vortex networks in the underlying TBG. Because of the magnifying effect of the topmost small-period graphene moiré, it is possible to obtain the tiny local lattice rotation by simply measuring the nanoscale periods of the topmost moiré pattern, which can be easily realized by using conventional microscopy techniques.

Figure 1a shows a schematic of a tiny-angle TBG with large period of moiré superlattice. The interfacial reconstruction is expected to occur by two rotation modes at the AA stacking regions and the AB (BA) stacking regions, shown as red and blue arrows respectively. In the past few years, many efforts have been made to measure the atomic reconstruction in the tiny-angle TBG. However, only very recently, the strain fields and the localized rotations of the TBG are measured by using a newly developed Bragg interferometry, based on four-dimensional scanning transmission electron microscopy [22]. In this work, we demonstrate that it is possible to direct and real-space visualize the local lattice rotations of the tiny-angle TBG by introducing another large-angle (small period) graphene moiré (see Fig. S1). By adding a graphene sheet on top of the tiny-angle TBG with a large twist angle, the studied system becomes twisted trilayer graphene (TTG) with a small period moiré generated between the first and the second layer $D_{12}$, and a large period moiré generated between the second and the third layer $D_{23}$, as schematically shown in Fig. 1a. The periods of moiré pattern $D$ generated between the topmost layer and the second layer at the AA and the AB (BA) stacking regions, labeled as $D_{12}^{AA}$ and $D_{12}^{AB}$ ($D_{12}^{BA}$) respectively, should be quite different because of the local lattice rotation of the tiny-angle TBG (Fig. 1a). We can obtain the local twist angles at the AA and the AB (BA) stacking regions, labeled as $\theta_{12}^{AA}$ and $\theta_{12}^{AB}$ ($\theta_{12}^{BA}$) respectively, based on the measured periods $D_{12}^{AA}$ and $D_{12}^{AB}$ ($D_{12}^{BA}$) according to $D = a/[2\sin(\theta/2)]$, where $a \approx 0.246$ nm (Fig. 1b) [23-27]. Then, the difference between the $\theta_{12}^{AA}$ and $\theta_{12}^{AB}$ ($\theta_{12}^{BA}$) directly reflects the local relative rotation between the AA and the AB (BA) stacking regions in the tiny-angle TBG.

To realize above measurements, the global twist angle between the topmost layer and the second layer, labeled as $\theta_{12}$, should be large enough to ensure that the period of moiré pattern generated between the topmost layer and the second layer is much smaller than the size of the AA stacking region in the tiny-angle TBG. Another advantage of the large twist angle is that it depends extremely sensitive on the period of moiré pattern (Fig. 1b), which ensures to measure the local lattice rotation in the tiny-angle TBG more accurately. In our experiment, the measurements were carried out on transferred TTG (see Fig. S2) on 0.7% Nb-doped SrTiO$_3$, as reported in Ref. [21], by using scanning tunneling microscopy/spectroscopy (STM/STS). In our TTG, the global twist angle between the second layer and the third layer, labeled as $\theta_{23}$, ranges from 0.24° to 1.6°, and the $\theta_{12}$ is usually larger than 2°. Figure 2 shows representative STM measurements of two TTG. Both samples can be identified by the characteristic "double-moiré" superlattices of the TTG[28], according to both the STM images (Figs. 2a and 2f) and their corresponding fast Fourier transforms (FFT) images (Figs. 2d, 2e, 2i and 2j). In the TTG, the large period moiré patterns shown in Figs. 2a and 2f are generated by the misorientations between the second and the third layer with $\theta_{23} \approx$ 1.01 ° (device D1) and 0.35 ° (device D2) respectively. In the tiny-angle TBG, even minor local heterostrain variations lead to large changes ($\sim 1/\theta$) in the moiré periodicity. Therefore, we observe more pronounced irregular moiré patterns in the device D2 TTG ($\theta_{23} \approx 0.35°$, Fig. 2f).

The structural reconstruction in the tiny-angle TBG results in large triangular Bernal (*AB* and *BA*) stacking domains and a triangular network of domain walls (DWs), which can host topological chiral conducting states [11,21,29,30]. Such a triangular network of conducting one-dimensional states is clearly observed in the STS maps of the device D2 TTG ($\theta_{23} \approx 0.35°$, see Supplementary material Fig. S3), demonstrating explicitly that there is interfacial reconstruction between the second and the third layer. Besides the "double-moiré" superlattices, the most pronounced feature observed in the STM images of the TTG is that the periods of moiré pattern generated between the topmost layer and the second layer $D_{12}$ vary at different positions (Figs. 2a and 2f). For example, $D_{12}^{AA} \approx 1.9$ nm and $D_{12}^{AB} \approx 2.1$ nm in the device D1 (Figs. 2b and 2c), which indicate $\theta_{12}^{AA} \approx 7.4$ ° and $\theta_{12}^{AB} \approx 6.7°$, suggesting a local lattice rotation between the AA and AB regions. Similarly, we obtained $D_{12}^{AA} \approx 4.0$ nm ($\theta_{12}^{AA} \approx 3.5°$) and $D_{12}^{AB} \approx 5.4$ nm ($\theta_{12}^{AB} \approx 2.6°$) in the device D2 (Figs. 2g and 2h). For the TBG with a large twist angle, the size of the moiré pattern is quite small. To minimize the intralayer elastic energy, the rigid TBG structure is energetically preferable and, hence, the periods of moiré pattern should be a constant. In the TTG shown in Fig. 2, the twist angle between the first and the second layer is quite large. Therefore, the observed variation of the periods of moiré pattern $D_{12}$ at different stacking regions is attributed to the local lattice rotation in the underlying small-angle TBG due to interfacial reconstruction, as illustrated in Fig. 1. We can directly obtain the local relative rotation between the AA and the AB (BA) stacking regions in the underlying TBG according to the difference between the $\theta_{12}^{AA}$ and $\theta_{12}^{AB}$ ($\theta_{12}^{BA}$).

To further explore effects of the local lattice rotation on the structure of the topmost TBG, we measured the periods of moiré pattern $D_{12}$ in the two exemplary TTG ($\theta_{23} \approx 1.01°$ and $0.35°$) as a function of positions, as shown in Figs. 3a and 3c respectively. Obviously, the measured results exhibit the same feature of moiré pattern in the underlying TBG of the TTG, which provides direct experimental evidence that the spatial variation of the $D_{12}$ arises from the structural reconstruction in the underlying TBG. According to $D = a/[2\sin(\theta/2)]$, the spatial distributions of the local rotation angle $\theta_{12}$ in the two exemplary TTG are plotted in Figs. 3b and 3d. Generally, the $\theta_{12}^{AA}$ is larger than the $\theta_{12}^{AB}$ ($\theta_{12}^{BA}$) and there is no obvious difference between the $\theta_{12}^{AB}$ and $\theta_{12}^{BA}$, which are consistent with the reconstruction mechanism predicted by theoretical studies[7-9]. Our method magnifies the local lattice rotation of the strain-field vortex networks by using moiré superlattice and allows us to directly image it with standard STM measurement. Actually, even the atomic resolution is unnecessary to measure the subtle local lattice rotation according to our experiment. Based on the results in Figs. 3b and 3d, we can also directly image the local relative lattice rotation in the triangular network of the DWs $\theta_{12}^{DW}$ and, generally, we obtain $\theta_{12}^{AA} > \theta_{12}^{DW} > \theta_{12}^{AB}$ ($\approx \theta_{12}^{BA}$).

Similar measurements were carried out in 15 TTG with different $\theta_{23}$, ranging from 0.24° to 1.6°, and their local lattice rotations are directly imaged by taking advantage of the magnified effect of moiré superlattice (see Supplementary material Figs. S5-S10 for more experimental data). Figure 3e summarizes the averaged local relative lattice rotation between different AA and AB regions, defined as $\varphi = \theta_{12}^{AA} - \theta_{12}^{AB}$, as a function of $\theta_{23}$ based on the measurement of the 15 TTG. As the $\theta_{23}$ nears zero, the $\varphi$ approaches a limiting value of approximately 1.0°. Extrapolation of $\varphi$ to large $\theta_{23}$ indicates that the onset of reconstruction in the TBG begins below about 1.7°.

The local lattice rotation in the underlying tiny-angle TBG not only leads to spatial variation of the twist angles, but also results in spatial modulated electronic properties of the topmost TBG. In the TBG, the twist angle determines period of the moiré pattern, which, consequently, determines the momentum separation of the two Dirac cones ($K_1$ and $K_2$), $\Delta K = 2|K|sin(\theta/2)$, between the adjacent graphene layers (here $|K| = 4\pi/3a$), as schematically shown in inset of Fig. 1b. A finite interlayer coupling $t_\theta$ generates a pair of saddle points with the energy separations that can be roughly estimated by the continuum model as $\hbar v_F \Delta K - 2t_\theta$ in the band structure of the TBG [4, 5, 23-27]. The two saddle points generate two pronounced VHSs with the energy separations $\Delta E_{VHS} \approx \hbar v_F \Delta K - 2t_\theta$ in the density of states (DOS), as shown in inset of Fig. 1b. Therefore, the spatial variation of the local twist angle $\theta_{12}$ induced by the underlying interfacial reconstruction in the TTG is expected to lead to spatial modulation of the $\Delta E_{VHS}$ in the topmost TBG. To explore this effect, we carry out scanning tunneling spectroscopy (STS) measurements in the TTG and two low-energy VHSs emerge in the STS spectrum as two pronounced peaks. The $\Delta E_{VHS}$ depends sensitively on the measured positions and, generally, the $\Delta E_{VHS}$ in the AA regions is larger than that measured in the AB and BA regions (Fig. 4a). Figure 4b shows the spatial variation of the $\Delta E_{VHS}$ measured in the device D2 (see Supplementary materials Fig. S11 for more experimental data on other TTG). Obviously, the spatial variation of the $\Delta E_{VHS}$ exhibit the same structure of the moiré pattern in the underlying TBG of the TTG, demonstrating explicitly that the structural reconstructions in the underlying TBG strongly modify the electronic properties of the topmost TBG. According to the result in Fig. 4b, there are other features that cannot be simply explained by the structural reconstruction in the underlying TBG. For example, it is surprisingly to find out that there is a slight difference between the values of $\Delta E_{VHS}$ in the AB and BA regions, as shown in top of Fig. 4b. The substrate can generate an effective electric field on the AB and BA regions and break their symmetry [29,30], which may slightly affect the values of $\Delta E_{VHS}$ in the two regions. Our experiment also indicates that the local heterostrain in the underlying TBG can affect the spatial variation of the $\Delta E_{VHS}$, as shown in bottom of Fig. 4b showing the region with a large anisotropy of the moiré periodicity. Such a result is reasonable because that the heterostrain of tiny-angle TBG can lead to a tetragonal structural transition in the moiré pattern, even resulting in a missing domain wall mode that separates the AB and BA regions [12,21].

In summary, by using the magnifying effect of the topmost small-period graphene moiré, we direct image the tiny local lattice rotation of small-angle TBG. The sub-Angstrom distortions of the lattice and tiny local rotation in the underlying small-angle TBG spatially modified twist angles and low-energy VHSs of the topmost small-period graphene moiré, which enable us to real-space image the strain-field vortex networks. Our results provide a general method to spatially engineer supported two-dimensional materials both in structures and electronic properties by using structure-reconstructed vdW systems as supporting substrates.

**Acknowledgments:**

This work was supported by the National Natural Science Foundation of China (Grant Nos. 11974050) and National Key R and D Program of China (Grant No. 2021YFA1400100). L.H. also acknowledges support from the National Program for Support of Top-notch Young Professionals, support from "the Fundamental Research Funds for the Central Universities", and support from "Chang Jiang Scholars Program".

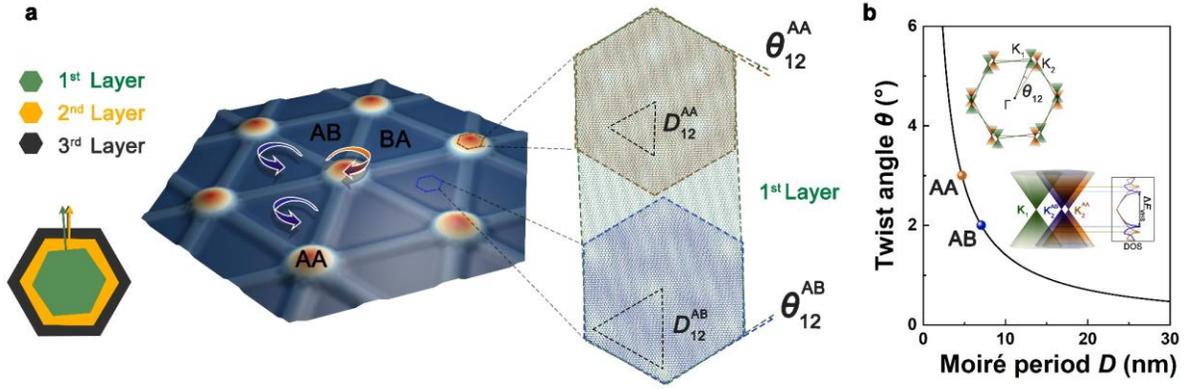

**Fig. 1.** Schematic view of local lattice rotation magnified by a topmost moiré. **a,** Left panel: The sketch of the TTG with top layer (green), second layer (yellow), and bottom layer (black). Middle panel: Schematic of the moiré pattern for underlying tiny-angle TBG. Right panel: Simulated images of local twist angles between the topmost layer and the second layer at the AA and the AB (BA) stacking regions of the underlying tiny-angle TBG. The difference between the $\theta_{12}^{AA}$ and $\theta_{12}^{AB}$ ($\theta_{12}^{BA}$) directly reflects the local lattice rotation between the AA and the AB (BA) stacking regions. **b,** Twist angle as a function of the moiré period according to $D = a/[2\sin(\theta/2)]$. The insets show the first Brillouin zone and the electronic band structure of TBG. $K_1$ and $K_2$ are the Dirac points of two layers. The overlap of the two Dirac cones, $K_1$ and $K_2^{AA}$ (or $K_2^{AB}$), generates two low-energy VHSs, which result in two peaks in the DOS.

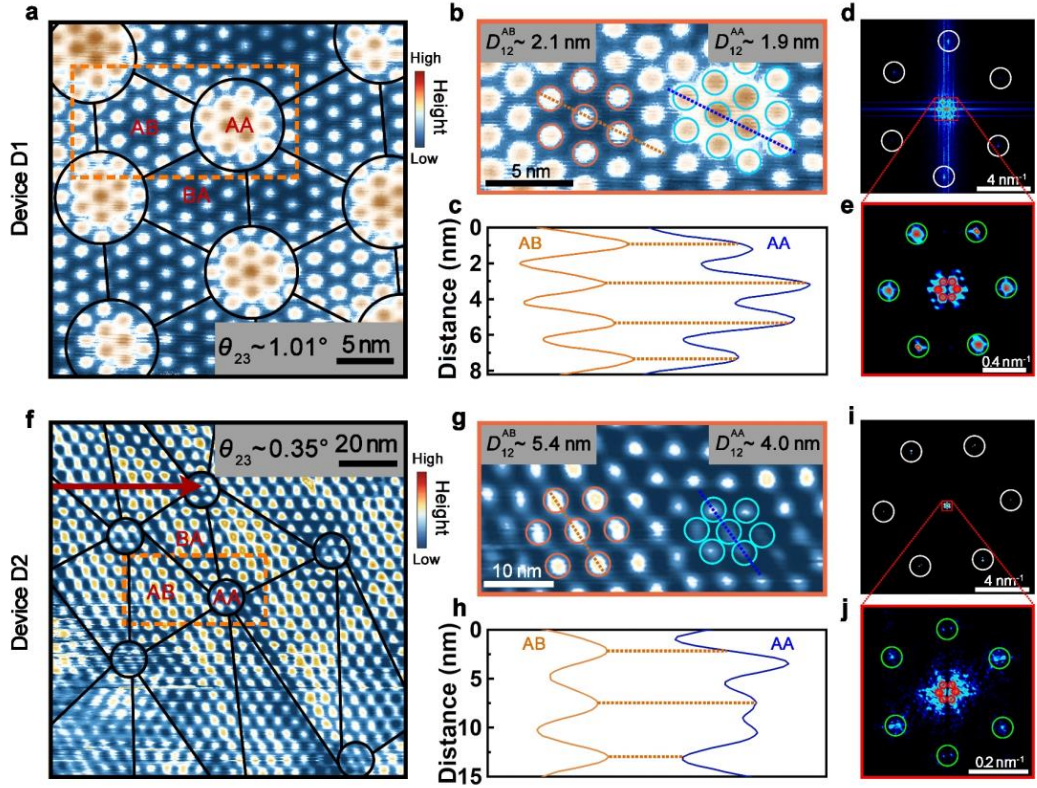

**Fig. 2.** Characterization of the TTG. **a** and **f**, STM topographies showing the "double-moiré" superlattices on $\theta_{23} \approx 1.01°$ and $\theta_{23} \approx 0.35°$ TTG samples. Topographies were taken at 1 V and 200 pA, and at 0.25 V and 400 pA, respectively. The overlaid circles and lines depict the moiré structure in the underlying TBG. **b** and **g**, STM images of the graphene area within the orange frame in **a** and **f**, respectively. The profile lines along the dashed lines are shown in **c** and **h**. The moiré periods of the topmost layer and the second layer $D_{12}$ vary at different positions. **d** and **e**, Corresponding FFT images of device D1. **i** and **j**, Corresponding FFT images of device D2. The white circles in **d** and **i** show clear reciprocal lattices of graphene. The green circles in **e** and **j** show the reciprocal moiré superlattices of the topmost two graphene layers, and the red circles in **e** and **j** show the reciprocal moiré superlattices of the underlying two graphene layers.

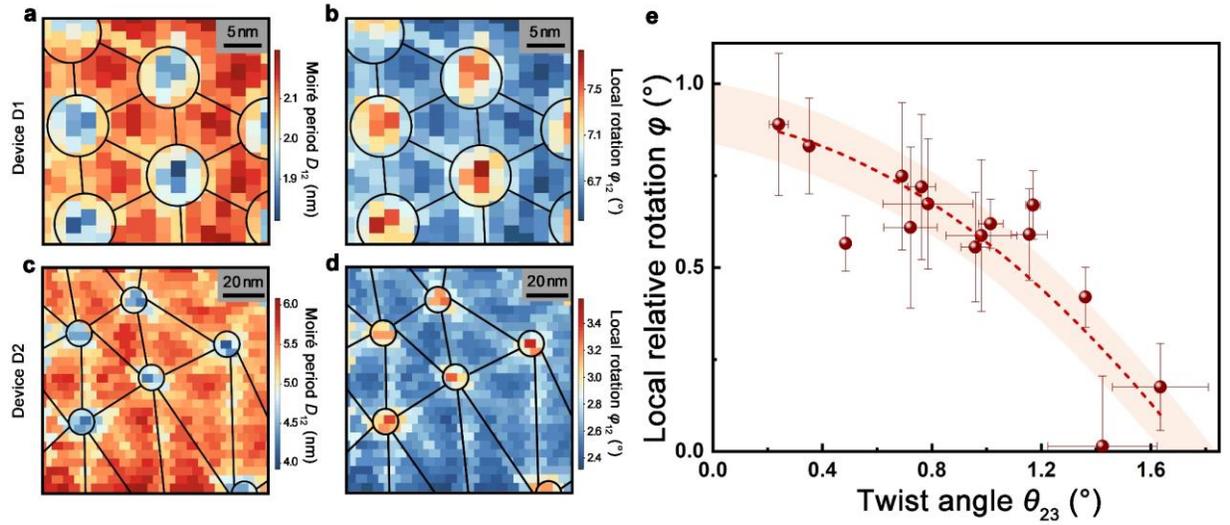

**Fig. 3.** Direct imaging the local lattice rotation magnified by a topmost moiré. **a** and **c**, The moiré period $D_{12}$ in the devices D1 and D2 as a function of positions. **b** and **d**, The spatial variations of the twist angle between the topmost layer and second layer. The difference of the twist angles between the AA and AB regions directly reflects the local lattice rotations in the underlying TBG. The overlaid circles and lines in **a-d** depict the moiré structure in the underlying TBG. **e**, The measured averaged local relative lattice rotation between the AA and AB regions ($\varphi = \theta_{12}^{AA} - \theta_{12}^{AB}$) as a function of $\theta_{23}$ on different devices. The error bars reflect the standard error of the data.

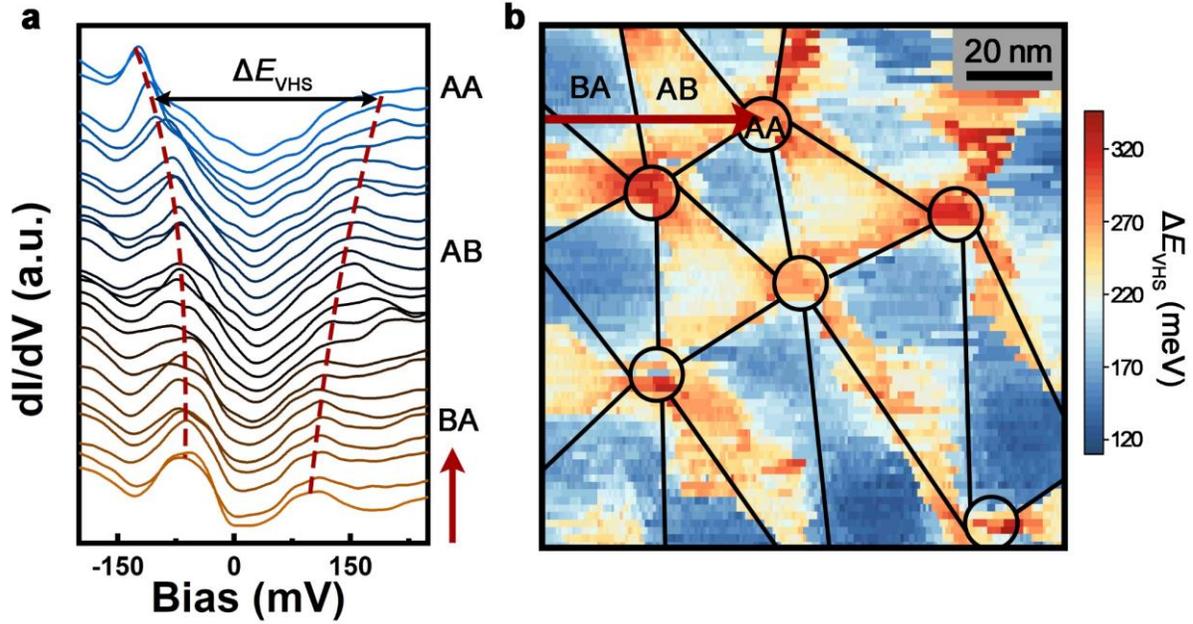

**Fig. 4.** Spatial variation of the low-energy VHSs induced by the local lattice rotation. **a**, The position-dependent d$I$/d$V$ spectra along the path indicated by the arrow in device D2 in Fig. 2f. **b**, Map of spatial variation of the $\Delta E_{VHS}$ in the topmost TBG, plotted according to 13160 spectra recorded at different positions. The overlaid circles and lines depict the moiré structure in the underlying TBG. The energy separations of VHSs $\Delta E_{VHS}$ in the AA region is generally larger than that in the AB and BA region due to the local lattice rotation.